\documentclass[11pt]{article}
\usepackage{bbm}

\title{\bf Singular compactifications and cosmology\thanks{Work supported by the `Schwerpunktprogramm Stringtheorie' of the DFG.}
\thanks{Talk given by T. Mohaupt at the EURESCO conference ``What comes beyond the Standard Model?'', 12. -- 17. July 2003, Portoro\v z, Slovenia.}
}

\author{ {\bf Laur J\"arv, Thomas Mohaupt and Frank Saueressig } \\ {\em Institute of Theoretical Physics, Friedrich-Schiller-University Jena, } \\ {\em Max-Wien-Platz 1, D-07743 Jena, Germany } \\ {\tt L.Jaerv, T.Mohaupt, F.Saueressig@tpi.uni-jena.de } }

\date{}

\begin{document}
\maketitle
\vspace{-0.5cm}
\begin{abstract}
\noindent We summarize our recent results of studying
five-dimensional Kasner cosmologies in a time-dependent Calabi-Yau
compactification of M-theory undergoing a topological flop
transition. The dynamics of the additional
states, which become massless at the transition point and give
rise to a scalar potential, helps to stabilize the moduli and
triggers short periods of accelerated cosmological expansion.
\end{abstract}
\vspace{0.5cm}

During the last year a lot of effort has been made to explain the astronomical evidence for an inflationary epoch of the early universe and the current modest accelerated expansion by invoking a scalar potential derived from string or M-theory compactifications. So far two
mechanisms leading to potentials viable to describe accelerated cosmological expansion have been explored
 \cite{Emp&Tow}: (i)
compactifications on hyperbolic spaces
\cite{hyper_comp}
and (ii) compactifications with fluxes \cite{flux_comp}. Our recent work \cite{model, cosmo} gives the first example of (iii) compactification on a singular internal manifold.

In the case of smooth compactifications one usually has a moduli space of vacua
corresponding to the deformations of the internal manifold $X$ and
the background fields. For theories with eight or less supercharges this moduli space includes special points where $X$
degenerates, rendering the corresponding low energy effective action (LEEA)
discontinuous or singular. However, within the full string or
M-theory these singularities are believed to be artifacts, which
result from ignoring some relevant modes of the theory, namely the
winding states of strings or branes around the cycles of $X$.
Singularities of $X$ arise when such cycles are contracted to zero
volume, thereby introducing additional massless states. 
Incorporating these states leads to a smooth {\em gauged} supergravity action which entails a scalar potential.

In Calabi-Yau (CY) compactifications of M-theory undergoing a topological flop transition these additional states (`transition states') are given by $N$ charged hypermultiplets which become massless at the transition locus. There are two ways to include the effect of these extra states in the LEEA. The usual LEEA is obtained by dimensional reduction on the smooth CY and contains
only states which are generically massless. The flop manifests itself in a discontinuous change
of the vector multiplet couplings at the transition locus. We call this
description the `Out-picture' since the extra states are left
out. On the contrary, the `In-picture' is obtained by including the
transition states as dynamical fields in the Lagrangian.

In \cite{model} we constructed an In-picture LEEA for a generic M-theory flop by combining knowledge about the general ${\cal N} = 2, D = 5$ gauged supergravity action with information about the extra
massless states.\footnote{This strategy was first applied in \cite{SU2} in the case of $SU(2)$ enhancement.}  While the vector multiplet sector could be treated exactly we used a toy model based on the quaternion-K\"ahler manifolds $\frac{U(1+N,2)}{U(1+N) \times U(2)}$ to describe the
hypermultiplet sector. In order to find the gauging describing the flop we
worked out the metrics, the Killing vectors, and the moment maps
of these spaces. This data enabled us to construct a unique LEEA which has all the properties to model a flop: the extra hypermultiplets acquire a mass away
from the transition locus while the universal hypermultiplet remains massless.

In~\cite{cosmo} we considered an explicit model for a CY compactification undergoing a flop with $N=1$ and investigated the effect of the transition states on five-dimensional
Kasner cosmologies,\footnote{This setup was previously considered in \cite{BL}, but there the hypermultiplet manifold was taken to be hyper-K\"ahler which is not consistent with local supersymmetry.} 
\begin{equation}
ds^2 = - d \tau^2 + e^{2 \alpha(\tau)} d \vec{x}^2 + e^{2 \beta(\tau)}
d y^2 \;.
\end{equation}
Comparing the cosmological solutions of the Out- and the In-picture, we found that the inclusion of the dynamical transition states  has  drastic consequences for 
moduli stabilization and accelerated expansion.

As soon as we allow all light states to be excited
the scalar fields no longer show the usual run-away
behavior but are attracted to the flop region
where they oscillate around the transition locus.
Thus the ``almost singular'' manifolds close to the flop 
are dynamically preferred. 
This is somewhat surprising, because the potential has still
many unlifted flat directions meaning there is no energy barrier
which prevents the system from running away.
Hence this effect cannot be predicted by just analyzing the critical
points of the superpotential.
The following thermodynamic analogy helps to explain the situation.
Generically the available energy of the system is
distributed equally among all the light modes
(``thermalization''). Thus near the flop line the additional
degrees of freedom get their natural share of it. Once this has
happened, it becomes very unlikely that the system ``finds'' the
flat directions and ``escapes'' from the flop region. Our numerical 
solutions confirm this picture:
irrespective of the initial conditions the system finally settles
down in a state where all the fields either approach
finite values or oscillate around the transition region with
comparable and small amplitudes. From time to time one sees
``fluctuations from equilibrium'', i.e., some mode picks up a
bigger share of the energy for a while, but the system eventually
thermalizes again. 
In
an ideal scenario of moduli stabilization, however, one would like to have a damped system
so that the moduli converge to fixed point values.

The second important aspect is that the scalar potential
of the In-picture induces short periods of accelerated
expansion in the three-space. Yet 
the net effect of the accelerating
periods on cosmic expansion is not very significant.
Again, this feature can be understood
qualitatively in terms of the properties of the scalar potential. 
The point is that the potential 
 is only flat along the unlifted
directions 
 while along the non-flat ones it is too steep to support sustained accelerated expansion. 
Transient periods of acceleration occur when the scalar fields pass through their collective turning point, where running ``uphill'' the potential turns into running ``downhill'' and the potential energy momentarily dominates over the kinetic energy.\footnote{This behavior is also common to the models of hyperbolic and flux compactifications, where likewise the acceleration is not pronounced enough for primordial inflation.}
To get
a considerable amount of inflation via a slow-roll mechanism, one would need to lift some of the flat
directions gently without making them too steep.

In summary we see that the dynamics of the transition states is
interesting and relevant, and can be part of the solution
of the problems of moduli stabilization and inflation. 
One direction for further investigations is
to consider more general gaugings of our five-dimensional model. 
Once gaugings leading to interesting cosmological solutions
are found, one should  clarify whether these can
be derived from string or M-theory where they correspond to adding fluxes or branes. 
Another direction is to extend our construction to other topological transitions.  In particular it would be interesting to study the effect of transition states on four-dimensional cosmologies arising, e.g., from type II compactifications on singular CY manifolds. It is conceivable that a realistic cosmology derived from sting or M-theory will have to include both the effects of fluxes and branes, and the possibility of internal manifolds becoming singular.

\end{document}